\documentclass[10pt,conference]{IEEEtran} 
\IEEEoverridecommandlockouts
\usepackage{tikz}
\usepackage{amsmath}
\usepackage{tikz}
\usepackage{amsmath}
\usepackage[version=4]{mhchem}
\usepackage{booktabs} 
\usepackage{graphicx}
\usepackage{microtype}
\usepackage{xcolor}
\usepackage{tcolorbox}
\usepackage{colortbl}
\usepackage{amsfonts}
\usepackage{fbox}
\usepackage{algorithm}
\usepackage{algpseudocode}
\usepackage{pifont}
\algrenewcommand\algorithmicensure{\textbf{Return:}}
\usepackage[]{collab}
\usepackage{textcomp}
\usepackage{booktabs}
\usepackage{tabularx}
\usepackage{ragged2e} 
\usepackage{balance}
\usepackage{xspace}
\usepackage{float}
\usepackage{multirow}
\usepackage{threeparttable}
\usepackage{enumitem}
\usepackage{tcolorbox}
\usepackage{graphicx}
\usepackage{hyperref}
\usepackage{url}
\usepackage{makecell}

\definecolor{background}{HTML}{D9E7FF}
\definecolor{edge}{HTML}{8686DD}
\newtcolorbox{mybox}{colback=background!30,
colframe=edge,
width=\columnwidth,
boxrule = 0.3mm,
top = 3pt, bottom=3pt, left=3pt, right=3pt
}

\newcommand{\ie}{{\em i.e.}\xspace}

\newcommand{\methodname}{UniSage\xspace}

\collabAuthor{zh}{magenta}{Zhihan}

\date{}

\author{
\IEEEauthorblockN{
Zhouruixing Zhu\IEEEauthorrefmark{1},
Zhihan Jiang\IEEEauthorrefmark{2},
Tianyi Yang\IEEEauthorrefmark{2},
and Pinjia He\IEEEauthorrefmark{1}\thanks{Pinjia He is the corresponding author.}
}
\IEEEauthorblockA{\IEEEauthorrefmark{1}
The Chinese University of Hong Kong, Shenzhen, China\\
zhouruixingzhu@link.cuhk.edu.cn, hepinjia@cuhk.edu.cn
}
\IEEEauthorblockA{\IEEEauthorrefmark{2}
The Chinese University of Hong Kong, Hong Kong, China\\
\{zhjiang22, tyyang\}@cse.cuhk.edu.hk
}
}

\usepackage[framemethod=TikZ]{mdframed}
\mdfdefinestyle{MyFrame}{%
    linecolor=black,
    outerlinewidth=0.3pt,
    roundcorner=5pt,
    skipabove = 5.5pt,
    skipbelow = 5.5pt,
    innertopmargin=5.5pt, 
    innerbottommargin=3pt,
    innerrightmargin=5.5pt,
    innerleftmargin=5.5pt,
    backgroundcolor=bg,
    nobreak=true, %
}

\begin{document}

\title{\methodname: A Unified and Diagnosis-Driven Sampling Framework for Microservice Systems}
\maketitle

\begin{abstract}

Traces and logs are essential observability signals for diagnosing microservice failures, but their massive volume makes full retention prohibitively expensive. Sampling is therefore widely used to reduce storage and analysis overhead. 
However, existing sampling frameworks suffer from two key limitations. First, they typically follow a \textit{sample-before-analysis} paradigm, where data is discarded before its diagnostic value is known, risking the loss of critical evidence for downstream analysis. Second, they mainly focus on trace sampling, while logs are either compressed or filtered independently, which can break the contextual linkage between anomalous traces and their explanatory log messages.

To address these limitations, we propose \methodname, a unified and diagnosis-driven sampling framework for microservice systems. \methodname is unified in two senses: it integrates lightweight analysis into the sampling pipeline to guide retention decisions, and it jointly samples traces and logs by exploiting their cross-modal correlations. 
Evaluations on three multi-modal microservice datasets show that \methodname consistently preserves more diagnostic evidence under tight budgets. At the 5\% budget, \methodname improves average diagnostic trace recall by 6.8\%--22.0\% over existing trace sampling baselines and achieves the highest diagnostic log recall on the evaluated log-labeled datasets. \methodname also remains practical, with runtime measurements showing modest overhead across the evaluated workloads.

\end{abstract}


\section{Introduction} \label{sec:intro}

Over the past decade, microservice architectures have become foundational to modern software systems by decomposing monoliths into loosely coupled, independently deployable services~\cite{Ali21Survey,huang2021sieve,steam}. This design improves elastic deployment and development agility, and has been widely adopted by industry companies including Amazon, Google, and Netflix~\cite{practice2017microservices}. 
The same decomposition, however, makes failures harder to diagnose. Site Reliability Engineers (SREs) must rely on observability data, including traces, logs, and metrics, to reconstruct system behavior~\cite{dapper,liu2023prism,logsurvey,jiang2024large,Hades}.

The volume of this telemetry makes full retention impractical. For instance, a large-scale e-commerce system at Alibaba generates approximately 20 petabytes of traces daily~\cite{mint}. Compression can reduce storage cost, but it does not resolve the diagnostic imbalance in observability data: most recorded behavior is normal and repetitive, while rare edge cases and failure indicators are often the evidence most useful for later troubleshooting. Sampling is therefore necessary, but its objective should be \emph{diagnostic utility}, not only fidelity to the dominant data distribution.

\begin{figure}[t]
    \centering
        {\includegraphics[width=1\linewidth]{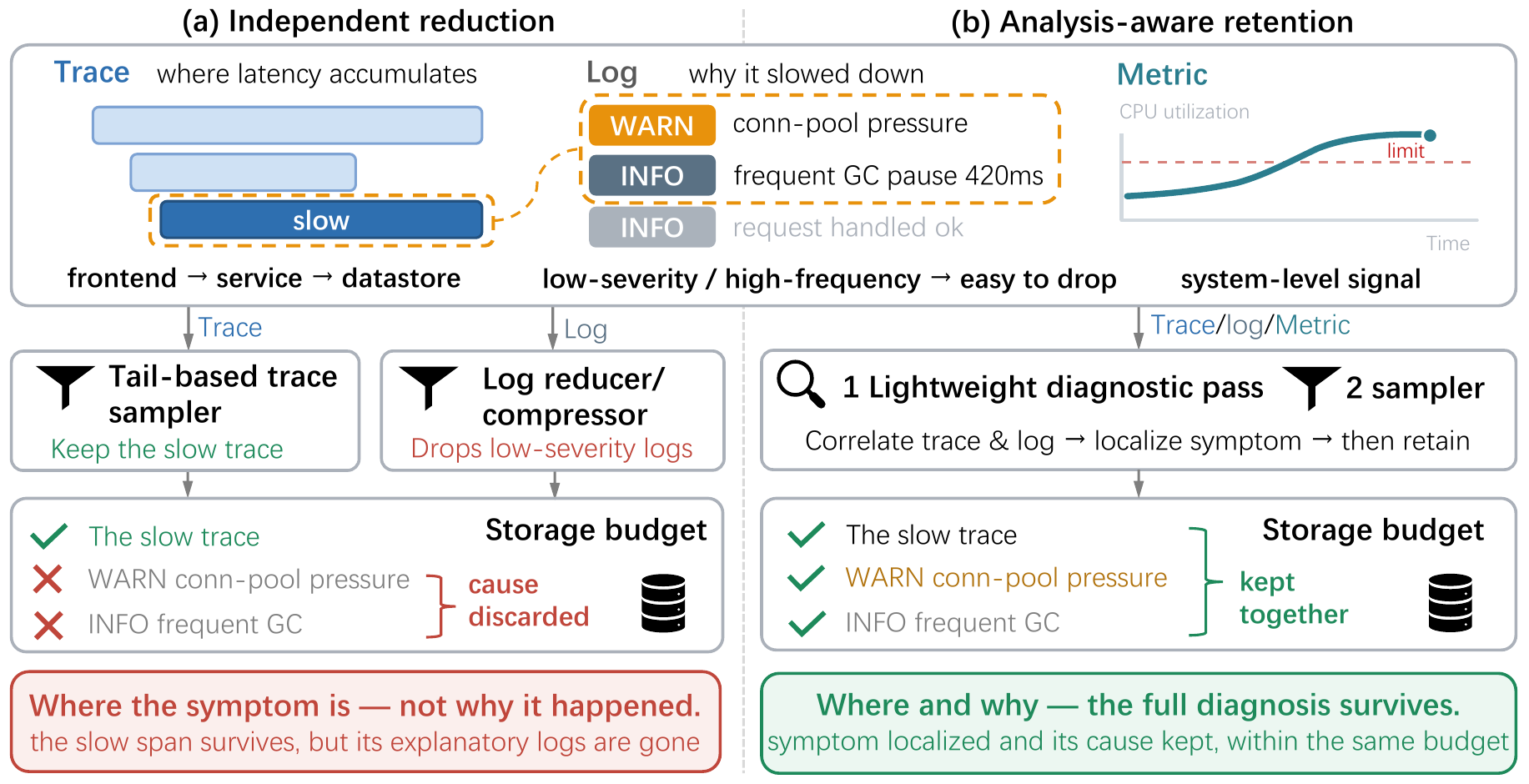}}
    \caption{
    Motivating example: independent reduction loses explanatory logs, while analysis-aware retention keeps trace–log evidence together.
    }
    \label{fig:example}
\end{figure}

Existing sampling methods usually decide what to retain before downstream diagnosis has established which evidence matters.
Tracing systems such as Jaeger~\cite{Jaeger} and Zipkin~\cite{Zipkin} commonly adopt uniform random sampling, a \textbf{head-based} strategy that decides whether to retain a trace before its execution outcome is observed.
\textbf{Tail-based} trace samplers decide after a trace is complete and can exploit richer trace information~\cite{huang2021sieve,sifter,TraStrainer,tracemesh,TracePicker}. However, they still follow a \textit{sample-before-analysis} workflow: telemetry is reduced before anomaly detection and root cause analysis, leaving only the retained subset for downstream diagnosis (Figure~\ref{fig:example}-a).
Recent log-enriched trace samplers such as Gleaner~\cite{yang2026gleanersemanticallyrichefficientonline} show that log context can improve trace selection, yet the retained unit remains centered on selected traces rather than diagnosis-guided retention of trace--log evidence.

This sample-before-analysis mismatch matters because diagnosis is not simply a search for unusual traces. A useful retained subset should preserve a coherent narrative: which request exposed the symptom, which service was implicated, and which surrounding events explain the failure. If the point at which retention is decided comes too early, the sampler may keep frequent normal paths because they dominate the stream, miss rare requests that later become relevant, or preserve a slow trace while discarding the logs that explain why it slowed down.

To address this \textbf{mismatch}, we advocate an \textit{analysis-first sampling paradigm} (Figure~\ref{fig:example}-b). Instead of deciding what to store before diagnosis, we first perform a transient diagnostic pass over the incoming telemetry and then write only the selected subset to long-term storage. This design does not persist full telemetry for later inspection; it targets the same storage and bandwidth bottleneck as conventional tail-based sampling while allowing diagnostic signals to influence retention. The shift is conceptual but important: a telemetry item is valuable when it helps preserve evidence for later diagnosis, not only when it statistically represents the full stream.

Realizing the analysis-first paradigm requires three properties. \textit{First}, the diagnostic pass must remain lightweight enough to run before retention; otherwise the sampler merely moves the bottleneck from storage to analysis. \textit{Second}, the guidance must be multi-modal because no single telemetry stream covers all fault manifestations. A trace can expose latency, a log can expose an exception or warning, and a metric can expose service-level resource pressure. \textit{Third}, retention must remain useful both when analysis identifies suspicious services and when the system appears normal, or the analysis signal is weak. A practical sampler therefore needs both diagnosis-driven relevance and an always-on coverage floor for rare-but-valid behavior.

We introduce \textbf{\methodname}, a unified and diagnosis-driven sampling framework for microservice systems. \methodname couples lightweight anomaly analysis with joint trace--log retention. Its dual-pillar strategy combines an always-on edge-case sampler, which preserves structurally or behaviorally rare telemetry, with an analysis-guided sampler, which prioritizes telemetry connected to diagnostic signals. A budget-aware retention step then converts these signals into retained traces and logs. In this way, we realized \textbf{unified} in two sense: analysis and sampling are placed in the same retention loop, and traces and logs are retained as correlated evidence rather than independent streams.

We evaluate \methodname on three multi-modal microservice datasets: Train Ticket~\cite{traintickt}, Online Boutique~\cite{OnlineBoutique}, and the real-world industrial AIOps21 dataset~\cite{Nezha,aiops21}. Across these datasets, \methodname consistently retains more diagnostically useful telemetry under tight storage budgets. At the 5\% budget, \methodname improves average diagnostic trace recall by 6.8\%--22.0\% over existing trace sampling baselines, while its log sampler achieves the highest diagnostic log recall on both datasets. 
These results validate the central premise of our analysis-first sampling: \textit{lightweight full-stream analysis can expose diagnostic signals that storage-limited sample-then-diagnose pipelines would otherwise discard.} 
Despite the additional analysis stage, \methodname remains practical: runtime experiments show that it keeps pace across the evaluated workloads, with execution time comparable to existing tail-based sampling baselines.
Finally, our robustness study confirms the value of \methodname’s dual-pillar design: when analysis signals become degraded, the always-on edge-case sampler provides a reliable fallback and prevents retention quality from collapsing into arbitrary or purely analysis-dependent sampling.

The key contributions of this paper are as follows:
\begin{itemize}[leftmargin=*, topsep=0pt, parsep=0pt]

\item \textbf{Analysis-first retention paradigm:} We formulate an \textit{analysis-first sampling paradigm} that lets lightweight diagnostic signals influence retention before telemetry is discarded, directly addressing the sample-before-analysis mismatch in conventional sampling pipelines.

\item \textbf{Unified trace--log retention:} We design and implement \methodname as a diagnosis-driven analysis--retention framework that keeps traces and logs as correlated evidence. Rather than treating modalities as independent streams, \methodname uses trace--log linkage to retain request paths together with explanatory runtime events.

\item \textbf{Dual-pillar budgeted sampling:} We combine analysis-guided relevance with always-on edge-case rarity, then convert these runtime signals into budgeted trace and log subsets. This design preserves a retention path for anomalous periods while maintaining coverage during normal or weak-signal periods.

\end{itemize}

\section{Background and Motivation}  \label{sec:motivation}

\subsection{Observability}

Observability, originally introduced in control theory~\cite{kalman1960general}, describes how internal system states can be inferred from external outputs. In modern microservice systems, this inference relies on traces, logs, and metrics~\cite{opentelemetry,ObservabilityBook}. Figure~\ref{fig:example} illustrates the roles used in this paper: traces organize request execution as spans and caller-callee edges; metrics summarize service health over time; and logs record runtime events, often with a \texttt{traceID} that links them to a request.
This linkage is central to retention. A trace can show where a symptom occurred, while logs can explain what happened inside the involved service. Metrics add service-level context and are usually compact enough to aggregate rather than sample~\cite{micrObser}. Logs and traces, by contrast, both grow quickly and are often reduced before troubleshooting, creating the risk that related evidence is no longer retained together~\cite{yu2024logCompression,logshrink,liu2019logzip}.

\subsection{Motivation}

\textbf{A diagnostic narrative can be broken by independent reduction.}
Consider a latency-sensitive service that suddenly slows down (Figure~\ref{fig:example}). A tail-based trace sampler may retain the slow request and expose the service where latency has accumulated. However, the root cause may be visible only in logs generated during the same request or time window, such as \texttt{WARN} messages about connection-pool pressure or \texttt{INFO} records indicating frequent garbage collection. If logs are reduced independently by severity, template frequency, or a separate budget, those explanatory messages may be discarded even though the symptomatic trace survives. The retained data then tells the operator where the symptom appeared, but not why it appeared.

This example illustrates the narrow but essential role of our motivation. The problem is not that traces, logs, or metrics are individually unhelpful; rather, their diagnostic value is often established only after symptoms are correlated and localized. A useful sampler should therefore let a lightweight diagnostic pass over the incoming stream to influence which trace--log evidence survives the storage budget, while still avoiding persistent full-stream retention.

This leads to three design requirements. First, retention should follow analysis so that diagnostic signals can be used before data is discarded. Second, retention should preserve correlated trace--log context, rather than sampling each modality as an isolated stream. Third, the budgeted subset should balance diagnosis-relevant evidence with rare-but-valid behavior, so the sampler remains useful during anomalous windows as well as normal or weak-signal periods.

\subsection{Problem Formulation} \label{sec:formulation}
We consider a microservice system $\mathcal{S}=\{s_1,\ldots,s_N\}$ of $N$ services with multi-modal telemetry. A fault-free construction phase $\mathcal{C}$ provides reference data $\mathbf{X}_{\mathcal{C}}=(\mathcal{T}_{\mathcal{C}},\mathcal{L}_{\mathcal{C}},\mathcal{M}_{\mathcal{C}})$, and the observation phase provides the incoming stream $\mathbf{X}_{\mathcal{O}}=(\mathcal{T},\mathcal{L},\mathcal{M})$, where $\mathcal{T}=\{t_i\}$, $\mathcal{L}=\{l_i\}$, and $\mathcal{M}=\{m_i\}$ denote traces, logs, and metrics. For each trace $t$, $\mathcal{S}_t$ denotes the services on its execution path, and the trace--log linkage $g(t)=(t,\mathcal{L}_t)$ maps $t$ to the logs carrying the same \texttt{traceID}.

\methodname follows an \emph{analysis-first} ordering: anomaly detection first predicts a system state $y\in\{0,1\}$; when $y=1$, RCA produces service-level scores $\Phi=\{\varphi_s\mid s\in\mathcal{S}\}$ and a ranked service list $\mathcal{R}$. We use \emph{diagnosis-driven} to describe the resulting retention objective: sampling should prioritize telemetry with expected diagnostic utility, not merely telemetry that is frequent or distributionally representative.

Under a per-modality retention budget $\beta$, let $K_\beta^{\mathcal{T}}=\lceil\beta|\mathcal{T}|\rceil$ and $K_\beta^{\mathcal{L}}=\lceil\beta|\mathcal{L}|\rceil$ denote the trace and log slots. For each telemetry item $x\in\mathcal{T}\cup\mathcal{L}$, \methodname computes an always-available edge-case value $p_e(x)$ independently of anomaly state or RCA output and, when $\Phi$ is available, an analysis-guided value $p_a(x)$. The final value is
\[
p(x)=
\begin{cases}
p_e(x), & y=0 \text{ or } \Phi \text{ is unavailable},\\
p_e(x)\cdot p_a(x), & y=1 \text{ and } \Phi \text{ is available}.
\end{cases}
\]
Here $p_a(x)$ denotes the post-correlation analysis-guided value after cross-modal propagation when that linkage is available. The sampling problem is to produce retained subsets $\mathcal{T}'_\beta\subseteq\mathcal{T}$ and $\mathcal{L}'_\beta\subseteq\mathcal{L}$ within these budgets while preserving diagnostic evidence and rare-but-valid behavior for downstream analysis.

\section{Methodology}

\subsection{Overview} \label{sec:overview}
\begin{figure*}[t]
    \centering
        {\includegraphics[width=0.9\linewidth]{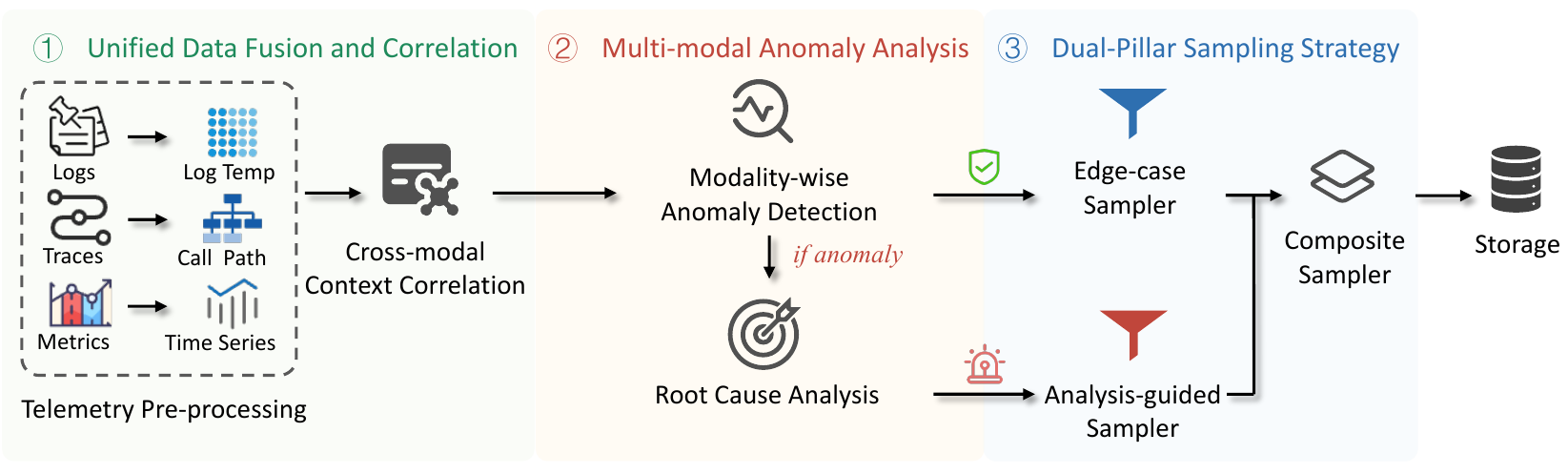}}
    \caption{Overview of \methodname.}
    \label{fig:overview}
\end{figure*}

Figure~\ref{fig:overview} presents \methodname, a unified \textit{diagnosis-driven sampling} framework for distributed microservice systems. Unlike samplers that discard telemetry before diagnosis, \methodname first extracts lightweight diagnostic signals from the incoming multi-modal stream and then decides which traces and logs to retain.
The framework has three phases. First, Unified Data Fusion and Correlation (Section~\ref{sec:data_fusion_correlation}) extracts modality-specific features and links traces with their logs. Second, Multi-Modal Anomaly Analysis (Section~\ref{sec:anomaly_analysis}) detects service-level deviations and, when anomalies are detected, scores suspicious services. Third, the Dual-Pillar Sampling Strategy (Section~\ref{sec:dual_pillar_sampling_strategy}) combines an always-on edge-case signal with an analysis-guided signal to produce the retained subset.
Across these phases, a telemetry item is valued by its expected diagnostic utility under a fixed retention budget $\beta$, rather than by its frequency in the dominant distribution. \methodname operationalizes this principle through edge-case rarity $p_e$, analysis-guided relevance $p_a$, and trace--log linkage $g$ (Section~\ref{sec:trace-log-correlation}); the final sampler composes these signals and applies a budget-aware retention rule.

\subsection{Unified Data Fusion and Correlation}
\label{sec:data_fusion_correlation}
The first phase converts heterogeneous telemetry into the compact representations used by anomaly analysis and sampling. It extracts trace, log, and metric features independently, then links traces and logs through their shared request identifier.

\subsubsection{Telemetry Pre-processing}
We apply modality-specific pre-processing to expose the signals used later by anomaly analysis and the edge-case sampler.

\textit{Trace Encoding.}
Traces provide two signals that are central to diagnosis-aware retention: span latency and invocation topology~\cite{huang2021sieve,TraStrainer}. We therefore encode each trace into span-level duration records and a set of unique root-to-leaf call paths $\mathcal{P}_t$. The former supports duration-based AD and RCA, while the latter supports topological rarity in the edge-case sampler.

\textit{Log Parsing.}
Logs are parsed into templates for automated analysis~\cite{jiang2024large,jiang2024lilac}. We employ Drain~\cite{drain}, a widely used log parser, to map unstructured log messages to parameterized templates. Each log entry is represented by its template, timestamp, service, and verbosity level.

\textit{Metric Aggregating.}
Metrics are aggregated into fixed-time windows to form multivariate service-state vectors. Metrics are used for analysis and service scoring, while long-term retention decisions in this work target traces and logs.

\subsubsection{Cross-Modal Context Correlation}
\label{sec:trace-log-correlation}
We assume a standard observability setup (\ie, OpenTelemetry~\cite{opentelemetry}) in which request logs carry the corresponding \texttt{traceID}. Using this identifier, we correlate the traces with their associated logs before analysis.
Formally, for each trace $t$, we define a correlation function $g$ that maps it to a contextual tuple $g: t \rightarrow (t, \mathcal{L}_t),$
where $\mathcal{L}_t$ denotes the set of log messages associated with trace $t$. This correlation supplies the request-level context used by cross-modal signal propagation (Section~\ref{sec:propagation}). When trace--log correlation is unavailable, \methodname disables propagation while keeping the remaining pipeline intact.

\subsection{Multi-modal Anomaly Analysis}
\label{sec:anomaly_analysis}

A core insight of the \methodname framework is its analysis-first sampling paradigm, where sampling decisions are guided by lightweight analysis over the \textit{full telemetry stream before retention}.
The analysis phase has two stages: modality-wise anomaly detection, followed by root cause analysis only when anomalies are detected. Its output is a service-level score used by the analysis-guided sampler (Section~\ref{sec:analysis-sampler}). This ordering does not require storing the full stream before sampling: the intended deployment performs feature extraction and scoring at the collector, then persists only the retained telemetry.

We use a deliberately narrow notion of \emph{anomaly}: a service-level statistical deviation from fault-free reference behavior, rather than a learned or semantic notion of failure. \emph{Lightweight} means that online scoring avoids deep models and per-incident training: log and trace detectors compare current signals with precomputed fault-free statistics, and metric detection uses a one-time PCA fit on fault-free reference data.

\subsubsection{Modality-wise Anomaly Detection}

The first stage detects anomalies independently for each modality because different fault types surface through different observability channels~\cite{Eadro,Hades}. Following evidence that simple statistical methods can be competitive in streaming and production settings~\cite{fang2025empirical}, we use $\kappa$-$\sigma$ checks for one-dimensional log and trace signals and PCA reconstruction error for metric vectors~\cite{pca-ae01}.

\begin{itemize}
    \item \textit{Log-wise Detection}: 
    The log detector monitors service-level shifts in failure-indicative log rates. Let $r^{\text{e}}$ and $r^{\text{w}}$ denote the rates of \texttt{ERROR} and \texttt{WARN} logs for a service, and let $r=\omega_{\ell}r^{\text{e}}+(1-\omega_{\ell})r^{\text{w}}$. Here $\omega_{\ell}$ is the ERROR-rate coefficient and $1-\omega_{\ell}$ the WARN-rate coefficient. The service is flagged when $r-\overline{r}>\kappa\sigma$, where $\overline{r}$ and $\sigma$ are estimated from fault-free reference data.

    \item \textit{Trace-wise Detection}:
    For traces, span duration is the primary signal of performance deviation. A service is flagged when the duration of its spans exceeds the fault-free mean by more than $\kappa$ standard deviations.
    \item \textit{Metric-wise Detection}: For metrics, we fit PCA on fault-free reference vectors and score each observation window by reconstruction error. If the error exceeds threshold $\rho$, the corresponding service and time window are flagged for root cause analysis.

\end{itemize}

\subsubsection{Root Cause Analysis}
\label{sec:root_cause_analysis}
When anomaly detection fires, \methodname performs root cause localization over collector-side features derived before retention. The goal is not to replace specialized offline RCA systems, but to produce a compact service ranking that can guide sampling while diagnostic evidence is still available.

\textit{Anomaly Service Scoring.}
\label{sec:anomaly_service_scoring}
For each modality $\mathcal{Z}\in\{\mathcal{T},\mathcal{L},\mathcal{M}\}$, the profiler emits service-associated deviation events $A_s^{\mathcal{Z}}$. Each event $a$ has a nonnegative deviation magnitude $\Delta_a^{\mathcal{Z}}$ and a severity rank $R_a^{\mathcal{Z}}$. The service score aggregates this evidence as
\[
\varphi_s =
\sum_{\mathcal{Z}\in\{\mathcal{T},\mathcal{L},\mathcal{M}\}}
\sum_{a\in A_s^{\mathcal{Z}}}
\frac{\Delta_a^{\mathcal{Z}}}{R_a^{\mathcal{Z}}+1},
\]
where higher-ranked events contribute more strongly. The resulting ranked service list is the input to the analysis-guided sampler (Section~\ref{sec:anomalous_service_bias}); after normalization, these scores define the analysis-guided value signal $p_a$.

\subsection{Dual-Pillar Sampling Strategy}
\label{sec:dual_pillar_sampling_strategy}
\begin{figure*}[t]
    \centering
        {\includegraphics[width=1\linewidth]{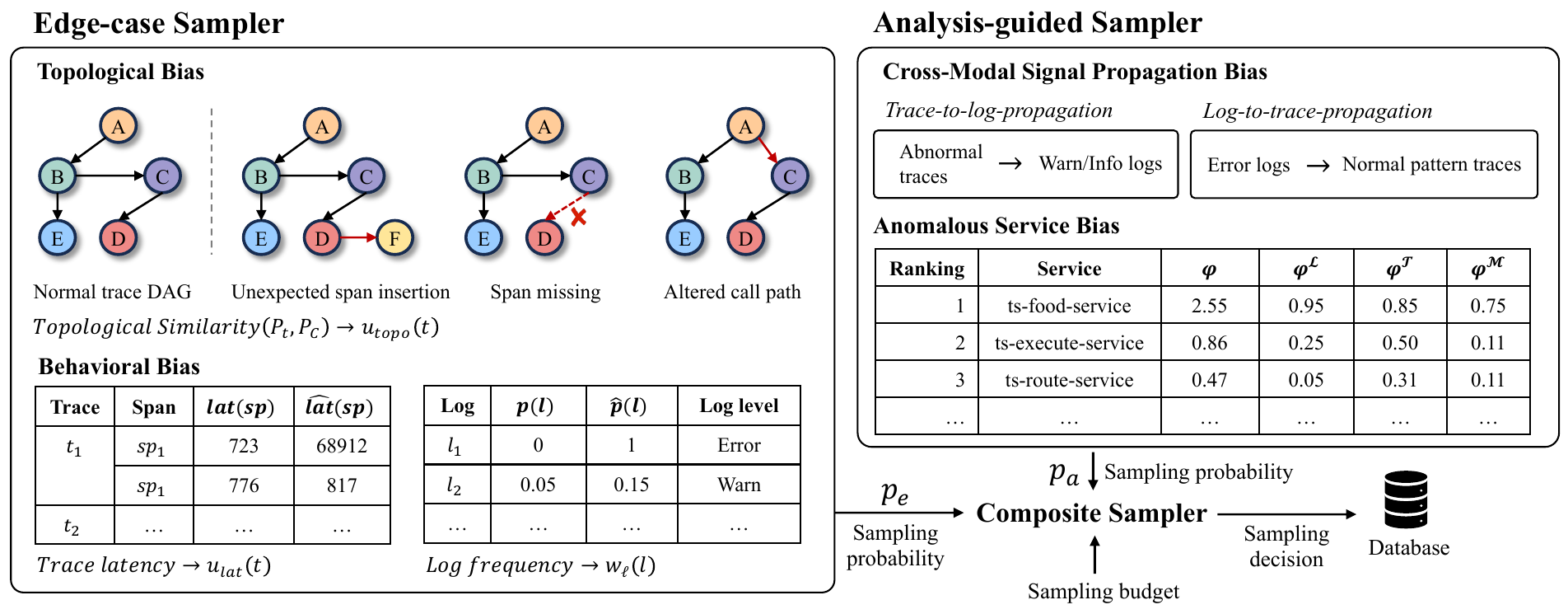}}
    \caption{Dual-Pillar Sampling Strategy of \methodname.
    }
    \label{fig:sample}
\end{figure*}

The final phase turns the diagnostic and rarity signals into sampling decisions. As illustrated in Figure~\ref{fig:sample}, \methodname uses two complementary pillars: an \textit{Edge-case Sampler}, which is always available and preserves rare behavior, and an \textit{Analysis-guided Sampler}, which becomes available when anomaly analysis yields an anomaly service ranking.
The composite sampler then converts these two signals into retained subsets under a fixed budget, using score-ranked retention as the default and opening a small diversity reserve only when the score prefix becomes structurally narrow.

\subsubsection{Edge-case Sampler}
The edge-case sampler assigns value to rare traces and logs independently of anomaly analysis. This pillar preserves novel or infrequent behavior even in normal windows and provides the fallback signal when the analysis pillar is silent.

\textit{Topological Bias.}
This bias captures structural deviations in distributed traces. A reference set of call paths $\mathcal{P}_{\mathcal{C}}$ is constructed from fault-free traces. For an incoming trace $t$, we compare its call paths $\mathcal{P}_t$ with the reference set using Jaccard similarity:
$\text{Sim}(c, c') = \frac{|c \cap c'|}{|c \cup c'|}.$
The normalized topological value is
\[
u_{\mathrm{topo}}(t)=\tanh\left(1-\max_{c\in\mathcal{P}_t,c'\in\mathcal{P}_{\mathcal{C}}}\text{Sim}(c,c')\right).
\]
Higher values indicate larger structural departure from the fault-free call-path reference.

\textit{Behavioral Bias.}
This bias evaluates statistical rarity in span latencies and log templates. For traces, we compute the maximum span-level z-score against the fault-free duration reference and map it through a monotone sigmoid to obtain $u_{\mathrm{lat}}(t)\in[0,1]$. The trace edge-case value then follows the implemented rule
\begin{equation}
p_e(t)=\max\left(u_{\mathrm{lat}}(t),u_{\mathrm{topo}}(t)\right).
\end{equation}
For logs, let $p_l'$ denote the reference probability of the template of log $l$, and let $w_{\ell}(l)$ denote its per-log verbosity weight. The log edge-case value is
\[
p_e(l)=w_{\ell}(l)\left(\frac{2}{1+e^{-2(-\log(p_l'+\epsilon))}}-1\right).
\]
Thus unseen and low-frequency templates receive high rarity values, while severity weighting prevents low-severity logs from dominating solely because they are rare. When anomaly detection flags nothing, or root-cause localization yields no usable service ranking, the analysis-guided signal is undefined and the composite sampler reduces to this edge-only fallback.

\subsubsection{Analysis-guided Sampler}
\label{sec:analysis-sampler}
The analysis-guided sampler is activated only when the analysis phase produces a usable service ranking. It raises the value of telemetry that traverses, originates from, or is linked to highly suspicious services.

\textit{Anomalous Service Bias.}
\label{sec:anomalous_service_bias}
Let $\tilde{\varphi}_s$ be the normalized RCA score for service $s$. For a trace $t$, \methodname uses the maximum score along its service path $\mathcal{S}_t$:
\(
p_a(t)=\tanh\left(\max_{s\in\mathcal{S}_t}\tilde{\varphi}_s\right).
\)
This max rule matches the single-root-service fault setting used in our evaluation and avoids assuming that symptoms must propagate uniformly to all downstream spans. For a log $l$ emitted by service $s_l$, the analysis-guided value is
\(
p_a(l)=w_{\ell}(l)\tanh(\tilde{\varphi}_{s_l}).
\)
The same verbosity weight used for log rarity keeps low-severity logs from dominating solely through service-level anomaly context.

\textit{Cross-Modal Signal Propagation Bias.}
\label{sec:propagation}
The trace--log correlation $g$ allows evidence in one modality to raise the analysis-guided value of linked evidence in the other. Propagation is guarded to avoid flooding the budget with weak or saturated inherited signals: only high-confidence sources can propagate, and a ceiling caps the inherited contribution. The guard never lowers an item's own score. For a source score $q$ and propagation strength $w$, let
\[
H(q,w)=\min(c,\ \mathbf{1}[q\ge\tau]\,(1-(1-q)^w)).
\]
Here $c$ is the propagation ceiling and $\tau$ the minimum source-score threshold.
For a log $l$ linked to trace $t_l$, and for a trace $t$ linked to logs $\mathcal{L}_t$, propagation updates the analysis-guided values with strengths $w_t$ for trace-to-log and $w_l$ for log-to-trace propagation as
\begin{equation}
    p_a'(l)=\max\big(p_a(l),H(p_a(t_l),w_t)\big),
\end{equation}
\begin{equation}
\begin{aligned}
    p_a^\star(t) &= \max_{l \in \mathcal{L}_t} p_a(l),\\
    p_a'(t) &= \max\big(p_a(t),H(p_a^\star(t),w_l)\big).
\end{aligned}
\end{equation}
For the composite rule below, $p_a$ denotes this propagated value when propagation is enabled.

\subsubsection{Composite Sampler}
\label{sec:budget_retention}
The composite sampler is the final decision layer of \methodname. It receives the always-on edge-case value $p_e$ and, when anomaly analysis produces a service ranking, the analysis-guided value $p_a$. For each trace or log item $x\in\mathcal{T}\cup\mathcal{L}$, it first forms a canonical score
\[
s(x)=
\begin{cases}
p_e(x), & \text{no analysis signal},\\
p_e(x)p_a(x), & \text{otherwise}.
\end{cases}
\]
The first branch is the operational edge-only fallback: \methodname still emits a ranking when no anomaly is detected or when root cause localization fails to produce a usable service order. The second branch requires both rarity and analysis relevance. Thus, a frequent fault log does not automatically consume the whole budget, and a rare but diagnostically irrelevant item does not outrank data tied to the suspected service. Max-based and weighted combinations are retained as ablation variants rather than the main composition.

\textit{Budget-aware selection.}
The composite score $s(x)$ provides the primary ordering, while the retention budget is applied separately to traces and logs:
\(
K_\beta^m=\max(1,\mathrm{round}(\beta |m|)),  m\in\{\mathcal{T},\mathcal{L}\}.
\)
For traces, \methodname first forms the score-ranked prefix $P_\beta^{\mathcal{T}}$. At very tight budgets, it returns this prefix directly. When the budget leaves room for diversity and the prefix is structurally narrow, a guarded reserve fills a small portion of slots from edge-ranked and previously uncovered structural patterns, using only sampler-observable parent--child span-edge keys from the current window. If the guard is not satisfied, the sampler falls back to $P_\beta^{\mathcal{T}}$. Logs use the same product composition over log anomaly relevance and service-template.

\section{Experimental Evaluation}

This section answers the following research questions (RQs):
\begin{itemize}[leftmargin=12pt, topsep=0pt]
\item \textbf{RQ1}: How effective is \methodname in retaining high-value (\ie, diagnostic and diverse) traces/logs?
\item \textbf{RQ2}: How does analysis-first sampling affect RCA utility under fair full-data and sampled-data diagnosis protocols?
\item \textbf{RQ3}: How much does each design component contribute?
\item \textbf{RQ4}: 
How robust is UniSage to degraded analysis signals and parameter perturbations, and what runtime overhead does it incur?
\end{itemize}

\begin{table*}[t]
\centering
\caption{Dataset statistics and evaluation scopes. M = million.}
\label{tab:dataset}
\begin{threeparttable}
\setlength{\tabcolsep}{4pt}
\renewcommand{\arraystretch}{1.05}
\begin{tabular}{@{}llrrrrrll@{}}
\toprule
Dataset & System & \#Cases & \#Services & \#Metrics & \#Traces & \#Logs (M) & Trace scope & Log scope \\
\midrule
Train Ticket & Railway Ticketing & 33 & 41 & 19 & 7,409 & 0.202 & problem-related + diversity & problem-related \\
Hipster & Online Boutique & 56 & 11 & 19 & 297,936 & 3.908 & problem-related + diversity & problem-related \\
AIOps21 & Industrial system & 159 & 14 & 483 & 747,125 & 1.325 & 123 primary + 36 diversity-only & diversity-only \\
\bottomrule
\end{tabular}
\end{threeparttable}
\end{table*}
\begin{table}[H]
\centering
\caption{Sampling metrics grouped by evaluation objective.}
\label{tab:evaluation_metrics}
\scriptsize
\setlength{\tabcolsep}{1pt}
\renewcommand{\arraystretch}{0.92}
\newcommand{\metriccheck}{\ding{51}}
\begin{tabularx}{\columnwidth}{@{}>{\RaggedRight\arraybackslash}Xccccc@{}}
\toprule
& \multicolumn{2}{c}{Modality} & \multicolumn{3}{c}{Dimension} \\
\cmidrule(lr){2-3}\cmidrule(lr){4-6}
Sampling Metric & Trace & Log & Diagnostic & Pattern & Fidelity \\
\midrule
Diagnostic Trace Recall (DTR) & \metriccheck & & \metriccheck & & \\
Diagnostic Trace Density (DTD) & \metriccheck & & \metriccheck & & \\
Trace-Pattern Coverage (TPC) & \metriccheck & & & \metriccheck & \\
Trace Distribution Fidelity (TDF) & \metriccheck & & & & \metriccheck \\
\midrule
Diagnostic Log Recall (DLR) & & \metriccheck & \metriccheck & & \\
Trace-Linked Evidence Recall (TLER) & & \metriccheck & \metriccheck & & \\
Trace-Log Diagnostic Pair Recall (TLDPR/Pair) & \metriccheck & \metriccheck & \metriccheck & & \\
Novel/Rare Log-Template Recall (NLTR/RLTR) & & \metriccheck & & \metriccheck & \metriccheck \\
Log-Template Coverage (LTC) & & \metriccheck & & \metriccheck & \\
Log Distribution Fidelity (LDF) & & \metriccheck & & & \metriccheck \\
\bottomrule
\end{tabularx}
\end{table}

\subsection{Experimental Settings}

\subsubsection{Datasets}
We evaluate \methodname on three multi-modal microservice datasets: Train Ticket~\cite{traintickt}, Hipster~\cite{OnlineBoutique}, and AIOps21~\cite{aiops21}. Train Ticket and Hipster come from the Nezha data collection~\cite{Nezha}; AIOps21 is an industrial challenge dataset from the International AIOps Challenge 2021.
Table~\ref{tab:dataset} summarizes the evaluated cases and modality scope. For AIOps21, 123 trace-observable cases (581,655 traces) enter problem-related trace-retention metrics, while 36 infrastructure cases (165,470 traces) are used only for diversity analyses. 
Problem-related log retention is reported only for Train Ticket and Hipster; AIOps21 logs are restricted to label-free diversity use because they lack line-level fault labels and reliable trace-linked \texttt{spanID} fields.
Across the three datasets, raw injection labels map to seven semantic fault types: CPU exhaustion/contention, memory pressure/leak, disk pressure, network delay, network packet-loss, code exception, and error return. Each case injects one fault into a single service; the fault can still produce cascading symptoms along invocation paths, so downstream callers may carry diagnostic trace or log evidence even though the injection target is unique.

\subsubsection{Annotation protocol}
We generate method-blind labels from raw telemetry, fault-free references, and injection metadata. Trace labels mark primary-scope traces that traverse the injected service (60 seconds after injection for Train Ticket/Hipster; the abnormal partition for AIOps21), while log labels mark primary-window symptoms (\texttt{ERROR}/\texttt{FATAL}, HTTP-5xx, or failure-related text). The resulting primary trace-label rates are 4.461\%, 8.058\%, and 25.244\% for Hipster, Train Ticket, and AIOps21, respectively. AIOps21 is higher because the label captures injected-service exposure rather than only visible symptoms.

\subsubsection{Evaluation metrics}
Table~\ref{tab:evaluation_metrics} groups the sampling metrics into three objectives. \emph{Diagnostic preservation} measures whether failure-relevant traces, logs, and trace-log links are retained. \emph{Behavioral coverage} measures whether the sample still spans rare trace structures and log templates. \emph{Distributional fidelity} measures whether retained telemetry remains close to the original distribution. We treat diagnostic preservation as the primary axis and use coverage and fidelity to expose the relevance-representativeness trade-off.
Stochastic sampling baselines are run with seeds 1--5, and the figures and tables report aggregate means from those runs.

\subsection{Baselines}

\subsubsection{Sampling baselines}
We compare \methodname with five SOTA trace samplers and \textit{Uniform} sampling. \textit{Uniform} represents head sampling and selects traces or logs with equal probability~\cite{tracemesh,TraStrainer,mint,steam}. \textit{Sieve}~\cite{huang2021sieve} uses robust random cut forest (RRCF)~\cite{RRCF} to prioritize structurally distinct traces. \textit{Sifter}~\cite{sifter} ranks traces by reconstruction loss from low-dimensional normal-path representations. \textit{TraStrainer}~\cite{TraStrainer} balances trace diversity with runtime state, \textit{TracePicker}~\cite{TracePicker} selects budgeted representative trace subsets, and \textit{Gleaner}~\cite{yang2026gleanersemanticallyrichefficientonline} enriches online trace sampling with associated log events.
Because most advanced baselines sample traces rather than logs, we evaluate log retention in two ways: direct \textit{Uniform} log sampling and trace-linked logs induced by each trace sampler. 
The latter is intentionally strong and fair: after a SOTA trace sampler selects traces, it retains all logs whose \texttt{traceID} values appear in those traces, and we align comparisons to the same effective log budget.

\subsubsection{RCA baselines}
We use three external RCA evaluators. \textit{TraceRCA}~\cite{TraceRCA} detects anomalous invocations against fault-free references and ranks services using trace evidence. 
\textit{MicroRCA}~\cite{microrca} localizes performance faults from service dependencies, invocation latency, and resource metrics.
\textit{Nezha}~\cite{Nezha} builds multimodal evidence from traces, logs, and metrics when cross-modal linkage is available. 
\methodname's built-in RCA (Section~\ref{sec:root_cause_analysis}) is used only as the upstream analysis signal for sampling and is reported separately as an analysis-first reference. More details of the implementation are introduced in Section~\ref{sec:rca_protocal}. 

\subsection{RQ1: Retention Quality}
\begin{figure*}[t]
    \centering
    \includegraphics[width=\textwidth]{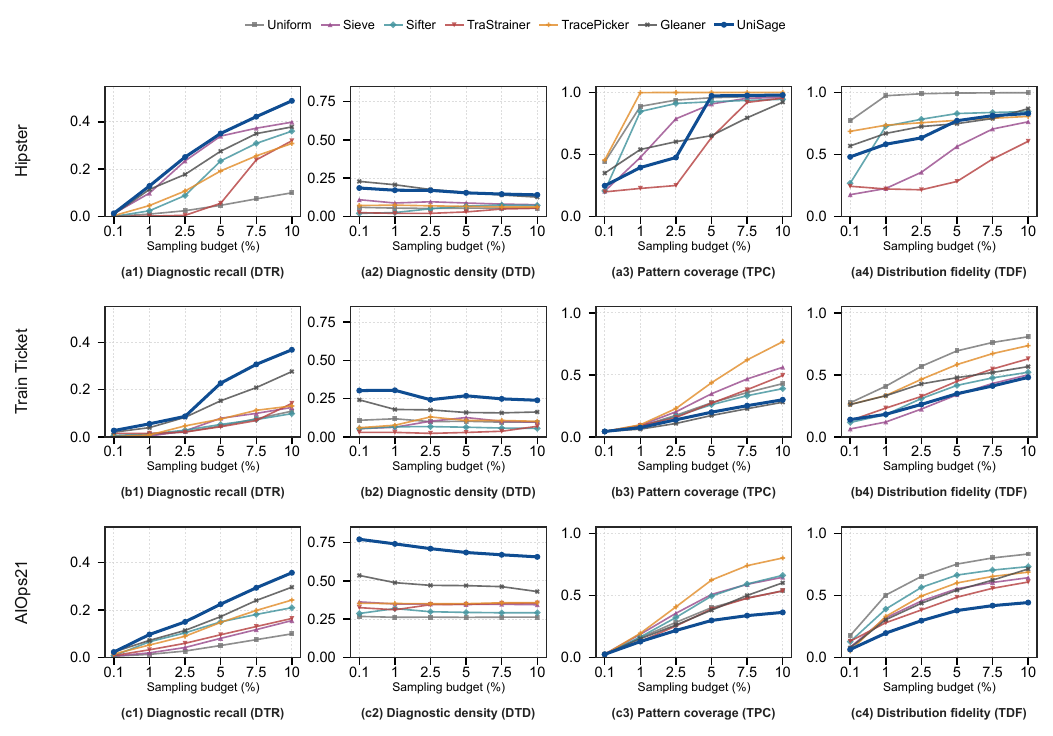}
    \caption{Trace retention across budgets. Rows are datasets and columns are trace-sampling metrics.}
    \label{fig:rq1_retention_summary}
    \label{fig:rq1_trace_budget_curves}
\end{figure*}

\begin{figure}[t]
    \centering
    \includegraphics[width=0.9\columnwidth]{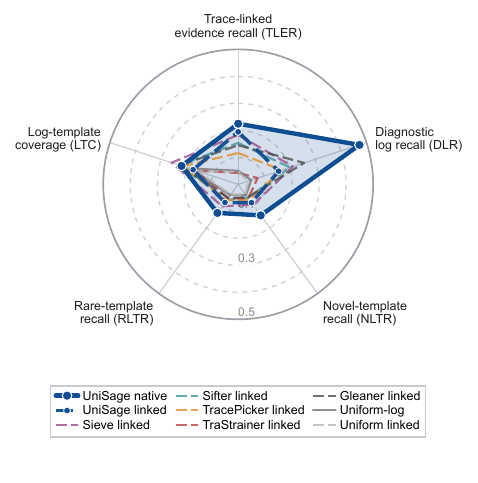}
    \caption{
    Log-retention profile at the 5\% budget, averaged over Hipster and Train Ticket.
    }
    \label{fig:rq1_log_budget_curves}
    \label{fig:rq1_log_radar}
\end{figure}

\begin{figure*}[t]
    \centering
    \includegraphics[width=0.85\textwidth]{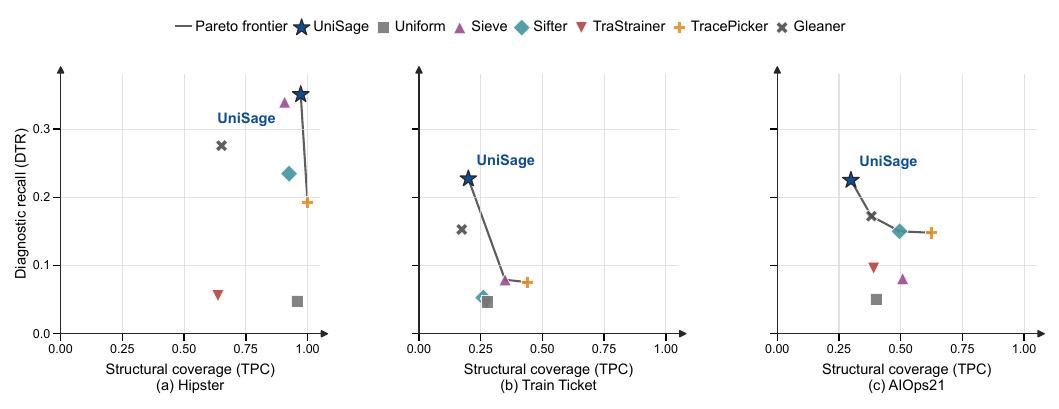}
    \caption{
    Diagnostic recall–pattern coverage trade-off at the 5\% budget.
    }
    \label{fig:rq1_dtr_tpc_tradeoff}
\end{figure*}

\subsubsection{Trace retention.}
Figure~\ref{fig:rq1_trace_budget_curves} shows that \methodname primarily improves diagnostic trace retention. Across six budgets and three datasets, it leads DTR in 17 of 18 dataset-budget panels; only AIOps21 at 0.1\% is marginally led by Gleaner. At the 5\% budget, \methodname achieves 0.351/0.227/0.225 DTR on Hipster/Train Ticket/AIOps21, compared with 0.340 for Sieve on Hipster and 0.153/0.172 for Gleaner on Train Ticket/AIOps21. It also leads DTD in 14 of 18 panels, indicating denser problem-exposing samples.

The TPC and TDF panels expose the corresponding relevance-representativeness trade-off. TracePicker has the strongest 5\% TPC, and Uniform has the highest 5\% TDF, while \methodname shifts budget toward diagnosis-relevant traces. Figure~\ref{fig:rq1_dtr_tpc_tradeoff} confirms that no baseline dominates \methodname in both DTR and TPC at 5\%: Hipster keeps near-maximal TPC (0.973 vs. TracePicker's 1.0), whereas Train Ticket and AIOps21 trade coverage for higher DTR.

\subsubsection{Log retention}
Figure~\ref{fig:rq1_log_radar} evaluates logs at the aligned effective-log 5\% budget, averaged over Hipster and Train Ticket. \methodname's native log sampler ranks first on TLER (0.225), DLR (0.472), NLTR (0.142), and RLTR (0.131). It exceeds the strongest non-native baselines by 0.216 DLR over Gleaner-linked logs and by 0.048/0.030 NLTR/RLTR over Sieve-linked logs, showing that trace-linked recovery misses some fault-manifesting log evidence.
Sieve-linked logs retain broader template coverage (LTC 0.259 vs. 0.221), so \methodname is not strongest on every coverage or fidelity axis. RQ1 therefore supports a narrower claim: under a fixed budget, \methodname prioritizes diagnostic trace-log evidence while maintaining enough diversity for follow-up analysis.

\subsection{RQ2: RCA Utility and Paradigm Comparison}
We evaluates RCA utility from two complementary views. The \textit{sample-then-diagnose} protocol tests whether the telemetry retained by each sampler remains useful to external RCA tools after storage reduction; fairness is enforced by using the same budget and the same evaluator within each comparison. The \textit{analysis-first} protocol tests \methodname's ordering choice by running RCA on full telemetry before retention, so diagnostic signals are computed before traces and logs are discarded. We therefore report the protocols separately and define their input contracts in Table~\ref{tab:rq2_rca_protocol}.

\textbf{RCA protocol.}
\label{sec:rca_protocal}
We evaluate RCA under two protocols. In \textit{analysis-first}, RCA runs on full telemetry before sampling, matching the signal available to \methodname during retention. In \textit{sample-then-diagnose}, each sampler first retains telemetry under the same budget, and the same external RCA evaluator then diagnoses the retained data. Table~\ref{tab:rq2_rca_protocol} separates the two fairness controls: the analysis-first comparison holds RCA input at full telemetry, while the sample-then-diagnose comparison holds the RCA evaluator and storage budget fixed and varies only the retained telemetry supplied by each sampler. It also defines the input contract for each evaluator; metrics remain full when required, while traces and logs follow each sampler's retained outputs.

\begin{table}[t]
\centering
\caption{
RCA input contracts for analysis-first/full-data and sample-then-diagnose protocols.
}
\label{tab:rq2_rca_protocol}
\scriptsize
\setlength{\tabcolsep}{1.8pt}
\renewcommand{\arraystretch}{1.12}
\begin{tabular}{l l l l l}
\toprule
\multirow{2}{*}{RCA evaluator} 
& \multirow{2}{*}{Modalities}
& \multirow{2}{*}{Full-data input}
& \multicolumn{2}{c}{Sample-then-diagnose input} \\
\cmidrule(lr){4-5}
& & & UniSage sampler & Baseline sampler \\
\midrule

TraceRCA 
& $\mathcal{T}$ 
& $\mathcal{T}$ 
& $\mathcal{T}^{U}_{\beta}$ 
& $\mathcal{T}^{B}_{\beta}$ \\

MicroRCA 
& $\mathcal{T},\mathcal{M}$ 
& $\mathcal{T},\mathcal{M}$ 
& $\mathcal{T}^{U}_{\beta},\mathcal{M}$ 
& $\mathcal{T}^{B}_{\beta},\mathcal{M}$ \\

Nezha 
& $\mathcal{T},\mathcal{L},\mathcal{M}$ 
& $\mathcal{T},\mathcal{L},\mathcal{M}$ 
& $\mathcal{T}^{U}_{\beta},\mathcal{L}^{U}_{\beta},\mathcal{M}$ 
& $\mathcal{T}^{B}_{\beta},\mathcal{L}^{B}_{\beta},\mathcal{M}$ \\

UniSage built-in RCA 
& $\mathcal{T},\mathcal{L},\mathcal{M}$ 
& $\mathcal{T},\mathcal{L},\mathcal{M}$ 
& -- 
& -- \\

\bottomrule
\end{tabular}
\vspace{0.1ex}
\begin{flushleft}
\footnotesize
$\mathcal{T}^{U/B}_{\beta}$ and $\mathcal{L}^{U/B}_{\beta}$ denote traces and logs retained by UniSage/baselines under budget $\beta$.
\end{flushleft}
\end{table}

\begin{table*}[t]
\centering
\caption{Downstream RCA under sample-then-diagnose, macro-averaged over the three datasets.}
\label{tab:rq2_downstream_rca}
\setlength{\tabcolsep}{1.4pt}
\begin{tabular}{l|cc|cc|cc|cc|cc|cc|cc|cc|cc}
\toprule
Sampling method & \multicolumn{6}{c|}{MicroRCA} & \multicolumn{6}{c|}{Nezha} & \multicolumn{6}{c}{TraceRCA} \\
\cmidrule(lr){2-7}\cmidrule(lr){8-13}\cmidrule(l){14-19}
 & \multicolumn{2}{c|}{$AC@1$} & \multicolumn{2}{c|}{$AC@3$} & \multicolumn{2}{c|}{MRR} & \multicolumn{2}{c|}{$AC@1$} & \multicolumn{2}{c|}{$AC@3$} & \multicolumn{2}{c|}{MRR} & \multicolumn{2}{c|}{$AC@1$} & \multicolumn{2}{c|}{$AC@3$} & \multicolumn{2}{c}{MRR} \\
 & 5\% & 10\% & 5\% & 10\% & 5\% & 10\% & 5\% & 10\% & 5\% & 10\% & 5\% & 10\% & 5\% & 10\% & 5\% & 10\% & 5\% & 10\% \\
\midrule
Gleaner & 0.18 & 0.20 & 0.37 & 0.35 & 0.33 & 0.33 & 0.25 & 0.33 & 0.39 & 0.42 & 0.37 & 0.42 & 0.14 & 0.15 & 0.31 & \textbf{0.31} & 0.27 & 0.26 \\
Sieve & 0.19 & 0.15 & 0.33 & 0.31 & 0.33 & 0.29 & 0.26 & 0.28 & 0.40 & 0.45 & 0.36 & 0.39 & 0.13 & 0.14 & 0.26 & 0.27 & 0.23 & 0.25 \\
Sifter & 0.12 & 0.14 & 0.31 & 0.35 & 0.28 & 0.29 & 0.22 & 0.28 & 0.35 & 0.44 & 0.30 & 0.37 & 0.08 & 0.11 & 0.16 & 0.22 & 0.14 & 0.18 \\
TracePicker & 0.17 & 0.18 & 0.32 & 0.37 & 0.31 & 0.32 & 0.28 & 0.34 & 0.40 & 0.46 & 0.37 & 0.44 & \textbf{0.17} & 0.12 & 0.29 & 0.25 & 0.26 & 0.23 \\
TraStrainer & 0.11 & 0.17 & 0.33 & 0.36 & 0.27 & 0.31 & 0.20 & 0.32 & 0.37 & 0.46 & 0.29 & 0.40 & 0.05 & 0.15 & 0.17 & \textbf{0.31} & 0.14 & 0.27 \\
Uniform & 0.15 & 0.16 & 0.33 & 0.32 & 0.30 & 0.30 & 0.25 & 0.26 & 0.38 & 0.43 & 0.34 & 0.37 & 0.16 & 0.15 & 0.22 & 0.24 & 0.22 & 0.23 \\
\cmidrule(l){1-19}
\methodname & \textbf{0.20} & \textbf{0.21} & \textbf{0.38} & \textbf{0.38} & \textbf{0.35} & \textbf{0.35} & \textbf{0.31} & \textbf{0.36} & \textbf{0.44} & \textbf{0.53} & \textbf{0.41} & \textbf{0.46} & \textbf{0.17} & \textbf{0.17} & \textbf{0.34} & \textbf{0.31} & \textbf{0.31} & \textbf{0.29} \\
\bottomrule
\end{tabular}
\end{table*}

\begin{table}[t]
\centering
\caption{
Full-telemetry RCA before sampling. Averages are computed over datasets.
}
\label{tab:rq2_analysis_first}
\scriptsize
\setlength{\tabcolsep}{1.4pt}
\renewcommand{\arraystretch}{1.0}
\begin{tabular}{@{}l|cc|cc|cc|cc@{}}
\toprule
\multirow{2}{*}{Dataset}
& \multicolumn{2}{c|}{MicroRCA}
& \multicolumn{2}{c|}{Nezha}
& \multicolumn{2}{c|}{TraceRCA}
& \multicolumn{2}{c}{\makecell[c]{\methodname\\built-in RCA}} \\
\cmidrule(lr){2-3}\cmidrule(lr){4-5}\cmidrule(lr){6-7}\cmidrule(lr){8-9}
& $AC@1$ & $AC@3$
& $AC@1$ & $AC@3$
& $AC@1$ & $AC@3$
& $AC@1$ & $AC@3$ \\
\midrule
Hipster & 0.29 & 0.46 & \textbf{0.93} & \textbf{0.93} & 0.29 & 0.45 & 0.75 & 0.89 \\
Train Ticket & 0.03 & 0.09 & \textbf{0.82} & \textbf{0.91} & 0.03 & 0.24 & 0.15 & 0.15 \\
AIOps21 & 0.20 & 0.49 & 0.15 & 0.20 & 0.05 & 0.05 & \textbf{0.45} & \textbf{0.62} \\
\midrule
Averages & 0.17 & 0.35 & \textbf{0.63} & \textbf{0.68} & 0.12 & 0.25 & 0.45 & 0.55 \\
\bottomrule
\end{tabular}
\end{table}

\textbf{Fair downstream RCA comparison.}
Table~\ref{tab:rq2_downstream_rca} reports the sampler-level comparison under sample-then-diagnose, macro-averaged over the three datasets. In this setting, every sampler is diagnosed by the same external RCA method after retention, so differences reflect the evidence each sampler preserves rather than the choice of RCA algorithm. Under both the 5\% headline budget and the 10\% sensitivity budget, \methodname achieves the best displayed macro-average on all three RCA metrics for MicroRCA, Nezha, and TraceRCA. Thus, even when judged by external RCA tools after sampling, \methodname provides more diagnosis-preserving inputs than sample-then-diagnose baselines.

\textbf{Analysis-first vs. sample-then-diagnose.}
Table~\ref{tab:rq2_analysis_first} reports the complementary analysis-first comparison, where RCA is performed on full telemetry before sampling. MicroRCA, Nezha, and TraceRCA serve as full-data external references, while \methodname's built-in RCA provides the lightweight in-situ analysis signal used by the sampler rather than serving as a downstream sampled-data evaluator. Nezha achieves the strongest external full-data RCA performance on average, but \methodname's RCA still reaches 0.45 $AC@1$ and 0.55 $AC@3$, exceeding the best sample-then-diagnose macro values in Table~\ref{tab:rq2_downstream_rca} (0.36 and 0.53 at 10\%). This suggests that even lightweight full-telemetry analysis can provide a stronger diagnostic signal than running RCA after storage sampling has already removed part of the evidence.
Taken together, Tables~\ref{tab:rq2_downstream_rca} and~\ref{tab:rq2_analysis_first} support the central ordering choice behind \methodname. Table~\ref{tab:rq2_downstream_rca} addresses the sampler fairness concern: under the same external RCA evaluator and storage budget, \methodname retains more useful telemetry for downstream diagnosis. Table~\ref{tab:rq2_analysis_first} addresses the full-data diagnosis concern: the analysis-first signal is not merely an artifact of \methodname's sampler, because it is evaluated alongside full-data external RCA references and remains competitive, while still being complementary to specialized full-data SOTA RCA methods.

\subsection{RQ3: Design Contributions}
\begin{table}[t]
\centering
\caption{Component effectiveness results. $\Delta$ are relative to full \methodname.}
\label{tab:rq3_ablation}
\scriptsize
\setlength{\tabcolsep}{8.5pt}
\begin{tabularx}{\columnwidth}{@{}>{\RaggedRight\arraybackslash}X|rrrrr@{}}
\toprule
Variant & $\Delta$DTR & $\Delta$TPC & $\Delta$TDF & $\Delta$DLR & $\Delta$Pair \\
\midrule
edge-only & -0.105 & -0.019 & +0.098 & +0.386 & +0.190 \\
analysis-only & -0.023 & -0.157 & +0.011 & -0.152 & -0.189 \\
max composition & -0.082 & -0.046 & +0.078 & +0.322 & +0.190 \\
weighted composition & +0.009 & -0.145 & -0.029 & +0.050 & +0.000 \\
product only & -0.005 & -0.140 & -0.016 & +0.000 & +0.000 \\
w/o log & -0.031 & +0.031 & +0.035 & +0.111 & -0.028 \\
w/o metric & -0.006 & +0.021 & +0.017 & +0.000 & +0.000 \\
\bottomrule
\end{tabularx}
\end{table}

Table~\ref{tab:rq3_ablation} decomposes \methodname's retention design. Removing the analysis-guided pillar (edge-only) reduces macro DTR by 0.105, confirming that edge rarity alone does not recover the same problem-related traces. Removing the edge pillar (analysis-only) has a smaller DTR loss (0.023) but reduces TPC by 0.157, showing that the edge sampler protects structural diversity and provides a fallback when analysis is weak or silent.
The modality ablations indicate targeted, dataset-sensitive contributions rather than uniform gains on every metric. Removing logs most clearly reduces Train Ticket diagnostic trace recall ($\Delta$DTR $=-0.091$ on Train Ticket; macro $-0.031$), while removing metrics has a smaller but concentrated effect on AIOps21 primary trace recall ($\Delta$DTR $=-0.017$; macro $-0.006$). Thus, log and metric signals complement the edge and analysis pillars by preserving different evidence channels under different workload conditions.

\subsection{RQ4: Robustness and Efficiency}
\begin{table}[t]
\centering
\caption{Robustness under corrupted RCA signals on AIOps21.}
\label{fig:rq4_robustness_curve}
\scriptsize
\setlength{\tabcolsep}{7pt}
\resizebox{\columnwidth}{!}{
\begin{tabular}{l|ccc}
\toprule
RCA corruption $\epsilon$ & Full \methodname & Analysis-only & Edge-only \\
\midrule
0.0 & 0.225 & 0.215 & 0.155 \\
0.5 & $0.180{\pm}0.011$ & $0.148{\pm}0.017$ & 0.155 \\
1.0 & $0.123{\pm}0.009$ & $0.061{\pm}0.010$ & 0.155 \\
\bottomrule
\end{tabular}
}
\end{table}

\begin{table}[t]
\centering
\caption{{Sensitivity study results. Values are macro-average changes from the default configuration.}}
\label{tab:rq4_sensitivity}
\scriptsize
\setlength{\tabcolsep}{1.8pt}
\begin{tabular}{@{}lcccc@{}}
\toprule
Knob & $\Delta$DTR & $\Delta$TPC & $\Delta$DLR & $\Delta$TLER \\
\midrule
$k$-$\sigma$ & $[-.004,+.000]$ & $[-.001,+.002]$ & $[-.005,0]$ & $0$ \\
PCA pct. & $0$ & $0$ & $0$ & $0$ \\
Log wt. & $[0,+.000]$ & $[0,+.000]$ & $[-.005,0]$ & $0$ \\
Prop. wt. & $0$ & $0$ & $[-.001,+.007]$ & $[-.018,0]$ \\
\bottomrule
\end{tabular}
\end{table}

\begin{figure*}[t]
    \centering
    \includegraphics[width=\textwidth]{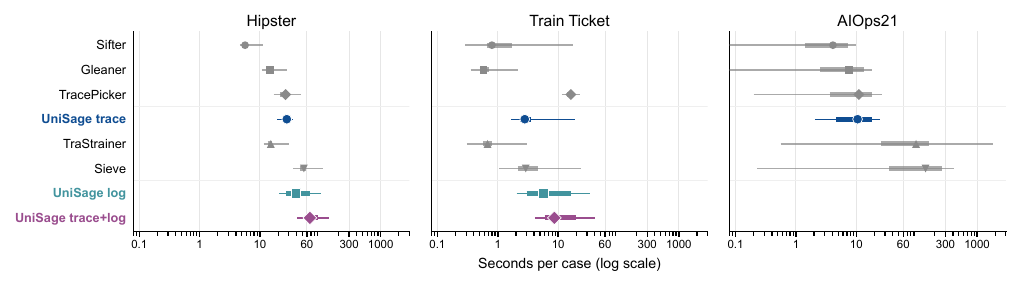}
    \caption{
    Per-case runtime distributions across datasets. Points denote medians; thick and thin bars denote IQRs and 5th--95th percentile ranges. \methodname includes analysis and sampling time; baselines include sampling time only.
    }
    \label{fig:rq5_runtime_envelope}

\end{figure*}

\subsubsection{Robustness to Analysis Quality}
Table~\ref{fig:rq4_robustness_curve} evaluates how retention changes when the upstream RCA signal degrades on AIOps21. We simulate degradation by randomly permuting service-level RCA scores for a fraction $\epsilon$ of services within each case, progressively decoupling the analysis signal from the true root cause.
Reducing observed RCA $AC@1$ from 0.593 to 0.340 lowers \methodname's DTR from 0.225 to 0.180, but \methodname still remains above the edge-only fallback of 0.155. Under extreme corruption ($\epsilon=1.0$), the analysis-only line collapses to 0.061, and the product-based \methodname falls to 0.123, while edge-only remains at 0.155. This is the intended safety interpretation: the edge pillar is the lower-bound fallback, and severe analysis failure should push the system toward that floor rather than arbitrary anomaly-adjacent retention.

\subsubsection{Hyperparameter sensitivity}
Table~\ref{tab:rq4_sensitivity} further tests whether \methodname's retention quality depends on finely tuned detector or propagation parameters. We vary the $\kappa$-$\sigma$ threshold ($\kappa \in \{2,3,4\}$), the PCA reconstruction-error percentile ($q \in \{85,90,95\}$), the log ERROR weighting ($0.9$, $0.7$, and $0.5$), and the trace--log propagation strength ($w_t,w_l \in \{0.25,0.5,0.75\}$) while keeping the remaining configuration locked. At the 5\% headline budget, the largest trace macro-DTR change is only $-0.004$ and the largest trace macro-TPC change is $+0.002$. For logs, the largest DLR change is $+0.007$, while overly strong propagation reduces TLER by $0.018$. The 10\% sanity budget shows the same pattern. These results indicate that the default configuration is robust under moderate perturbations rather than a finely tuned point solution.

\subsubsection{Runtime Overhead}
Figure~\ref{fig:rq5_runtime_envelope} completes RQ4 by reporting per-case elapsed processing time as median and quantile envelopes. We use distributional summaries because the measurements are skewed by case size and I/O long tails. The timings characterize the current batch implementation. For \methodname, the measured trace route invokes anomaly detection and root-cause localization before sampling, so the reported time includes the pre-sampling analysis stage rather than only the final ranking step. The trace+log route is a derived sum of separately measured trace and native-log routes, not a fused production timer.

The three datasets stress different costs. Hipster has the largest log workload (3.91M logs, median 74.3K logs per case) and many per-case log templates, so native log retention visibly increases processing time; even then, \methodname's trace+log median is 67.1 seconds. Train Ticket is much smaller in volume (7,409 traces and 0.20M logs) but has more structural and template sparsity: 13.7\% of traces fall into the rare-structure stratum and 61.4\% of logs fall into the rare-template stratum. Its \methodname trace+log median is therefore only 8.7 seconds. AIOps21 has the highest metric dimensionality (483 metrics) and the largest trace scope, yet \methodname's trace-route median is 10.5 seconds, close to TracePicker (11.0 seconds) and Gleaner (7.6 seconds), and far below the long-tail medians of Sieve (140.3 seconds) and TraStrainer (97.7 seconds).

The processing time is also below the observed telemetry window of each case. The median case windows are about 3.0 minutes for Hipster, 3.0 minutes for Train Ticket, and 5.0 minutes for AIOps21, while \methodname's median processing times are 27.8 seconds for Hipster trace sampling, 67.1 seconds for Hipster trace+log retention, 2.8/8.7 seconds for Train Ticket trace/trace+log retention, and 10.5 seconds for AIOps21 trace sampling. Thus, even after including the pre-sampling RCA/RCL analysis, \methodname remains in the same practical runtime range as strong trace baselines and can keep pace at case granularity on the evaluated workloads. We interpret this as evidence of deployability of the current pipeline, not as a fully optimized streaming-system benchmark.

\section{Discussion}

\subsection{Limitations}

\methodname adopts a tail sampling strategy, which is inherently biased toward informative samples and therefore incurs higher computational overhead than head sampling methods based on random selection. 
Nevertheless, our overhead remains fully acceptable, demonstrating the practicality of \methodname.
An interesting direction for future work is to explore hybrid sampling strategies that integrate head and tail sampling, enabling a principled trade-off between efficiency and effectiveness.
The analyzer itself is intentionally minimal: the $\kappa$-$\sigma$ rule and PCA keep the online analysis stage simple and help isolate the contribution of analysis-guided retention rather than the sophistication of a particular detector. Studying stronger or adaptive analyzers remains future work.

\subsection{Scope and Generalizability}
Although our evaluation centers on RCA, \methodname is not a failure-only filter. Its edge-case sampler (Section~\ref{sec:dual_pillar_sampling_strategy}) is \emph{always active} and independent of the anomaly verdict, retaining structurally rare call paths and behaviorally rare spans such as latency outliers even when no fault is detected. This coverage floor supports tasks such as program understanding and performance analysis, while the analysis-guided sampler adds fault-relevant retention when diagnostic signals appear. The same analysis-first principle could also guide adjacent reduction mechanisms, such as compressing normal log windows more aggressively, retaining finer-grained metrics for RCA-implicated services, or biasing head sampling with collector-side anomaly summaries. We therefore view trace and log sampling as the concrete instance evaluated in this paper rather than the boundary of the methodology.

\subsection{Threats to Validity}
The main threat to \textit{internal validity} lies in the implementation of baselines. For Sieve and TraStrainer, we use their publicly available open-source implementations. However, Sifter lacks open-source code, so we carefully re-implement it based on descriptions in the original paper, ensuring the use of identical libraries and settings whenever possible.
Threats to \textit{external validity} still stem from dataset coverage: three datasets cannot represent all production failure modes or all deployment scales. However, the evaluation is not narrow along a single axis. Train Ticket contributes the largest service graph in our study and, among the open-source benchmarks, a higher rare-structure stratum. Hipster stresses log volume and template diversity, with 3.908M logs and many per-case log templates. AIOps21 contributes industrial telemetry, the most cases and traces, and the highest metric dimensionality (483 metrics), although its service topology is smaller than some enterprise deployments. Together, these workloads exercise service scale, log diversity, trace volume, and metric dimensionality. Generalization to broader deployment scales and settings remains future work, as does a fused streaming deployment with per-stage timing and peak-memory instrumentation.

\section{Conclusion}
In this paper, we introduced \methodname, a unified sampling framework for preserving diagnostic utility under tight observability budgets. \methodname shifts sampling from a sample-before-analysis pipeline to an analysis-first sampling strategy, using lightweight anomaly detection and root cause analysis to retain traces and logs whose value may only become clear after diagnosis.
Across three trace-observable datasets and two log-labeled datasets, \methodname improves diagnostic retention while maintaining competitive downstream RCA utility and practical route-level overhead on the evaluated workloads. These results support its dual-pillar design: analysis-guided sampling preserves fault evidence, while the always-on edge sampler maintains diversity and robustness when analysis signals are weak. Overall, \methodname shows that diagnostic context can serve directly as a retention signal for more useful observability reduction.

\section{Data Availability}
We make our code and datasets publicly available at \url{https://github.com/zhouruixingzhu/UniSage_Sampling}.

\section{Acknowledgments}
This paper was supported by the Guangdong Basic and Applied Basic Research Foundation (No. 2024A1515010145) and the Shenzhen
Science and Technology Program (Shenzhen Key Laboratory Grant No. ZDSYS20230626091302006).

\IEEEtriggeratref{20}
\IEEEtriggercmd{\balance}

\bibliographystyle{IEEEtran}
\bibliography{reference}

\end{document}